%% file: paper25.tex
\documentclass[10pt,conference,compsocconf,letterpaper]{IEEEtran}
\usepackage[pdftex]{hyperref} 
\usepackage{setspace}
\usepackage{amssymb}
\usepackage{amsmath}
\usepackage{algorithmic}
\usepackage{algorithm}
\usepackage{graphicx}
\usepackage{epsfig}
\usepackage{graphics}
\usepackage{multicol}
\newtheorem{theorem}{Theorem}[section]
\newtheorem{lemma}[theorem]{Lemma}
\newtheorem{corollary}[theorem]{Corollary}
\input{macros}

\usepackage{color}

\newcommand{\fakeparagraph}[1]{\vspace{1.8mm}\noindent\textbf{#1}}

\begin{document}
\title{Order-Preserving Renaming in Synchronous Systems with Byzantine Faults}
\author{
{Oksana Denysyuk and Lu\'{i}s Rodrigues}%
\vspace{1.6mm}\\
\fontsize{10}{10}\selectfont\itshape
INESC-ID, Instituto Superior T\'{e}cnico, Universidade T\'{e}cnica de Lisboa, Portugal\\
\fontsize{9}{9}\selectfont\ttfamily\upshape
\{oksana.denysyuk,ler\}@ist.utl.pt
}
\maketitle
\begin{abstract}
Renaming is a fundamental problem in distributed computing, which consists of a set of processes picking distinct names from a given namespace. The paper presents algorithms that solve order-preserving renaming in synchronous message passing systems with Byzantine processes. To the best of our knowledge, this work is the first to address order-preserving renaming in the given model.  Although this problem can be solved by using consensus, it is known that renaming is ``weaker" than consensus, therefore we are mainly concerned with the efficiency of performing renaming and make three contributions in this direction. We present an order-preserving renaming algorithm for $N > 3t$ with target namespace of size $N+t-1$ and logarithmic step complexity (where $N$ is the number of processes and $t$ is an upper bound on the number of faults). Similarly to the existing crash-tolerant solution, our algorithm employs the ideas from the approximate agreement problem. We show that our algorithm has constant step complexity if $N>t^2+2t$ and achieves tight namespace of size $N$. Finally, we present an algorithm that solves order-preserving renaming in just $2$ communication steps, if $N > 2t^2 + t$.
\end{abstract} \thispagestyle{plain}
\section{Introduction}\label{sec:intro}
Renaming is a fundamental problem in distributed computing, which can be informally described as follows: a set of processes $\{p_1, \cdots, p_N\}$ with unique ids in the range $[1 \cdots N_{max}]$ must pick new names from a given range $[1, \cdots, M]$, where $M \ll N_{max}$. The range of values to which new names belong is called \emph{target namespace}. In this work, we are interested in an order-preserving variant of the renaming problem that requires processes to preserve the order of their old names. This variant is interesting as it allows to use renaming in settings where the original identifiers encode some additional information, such as, for instance, their relative priority in accessing a shared resource.

In this paper we address this problem in synchronous message-passing systems subject to Byzantine faults. In synchronous systems, order-preserving renaming has only been previously addressed for crash-faults\cite{order-preserving-synch}. Adapting previous work to cope with Byzantine processes raises several interesting challenges. First, Byzantine nodes may lie about their ids, use different ids when communicating with different processes, and $t$ faulty processes may even collude to create more than $t$ ids, none of which can be identified as bogus by correct processes. Secondly, Byzantine processes can lie about the ids they have seen, sending contradictory information to different correct processes. This breaks the algorithms designed for crash-faults\cite{order-preserving-synch} in different ways. Interestingly, some of the most ``intuitive'' approaches to tackle these challenges reveal themselves inadequate:

$\bullet$ One could consider using Reliable Broadcast\cite{BrachaToueg} or consensus\cite{consensus} to ensure each process agree on the same set of identifiers and, in this way, solve renaming, but these approaches have step complexity linear in the number of faults\cite{consensus}.

$\bullet$ There are techniques to translate a crash-tolerant algorithm into a Byzantine-tolerant algorithm\cite{crash-to-byz1,crash-to-byz2}, but they have two main limitations. First, they blow up the message and step complexity because processes must broadcast (and echo) histories of previously received messages. Second, these techniques assume that, when a process receives a message, it knows the id of the sender. But, with this knowledge it becomes trivial to solve the order-preserving renaming problem without any communication (just by sorting the set of ids and then choosing the rank of each id as new name).

$\bullet$ Finally, the crash-tolerant algorithm by Okun\cite{order-preserving-synch} is based on running multiple instances of Approximate Agreement (AA) to agree on a rank for each id. We could consider that a Byzantine-tolerant version of that algorithm could be easily obtained just by replacing AA in~\cite{order-preserving-synch} by some Byzantine-tolerant version of AA, such as~\cite{approx-agreement}. In fact, Byzantine-tolerant AA guarantees that the outputs are within the range of values issued by the correct processes. Unfortunately, Byzantine nodes can cause correct processes to propose overlapping intervals of values for different instances of AA and, therefore, the outputs may not preserve the initial ordering.

This paper takes on the latter idea of using Byzantine-tolerant approximate agreement to solve order-preserving renaming. For this purpose, we introduce a 4-step id selection scheme that restricts the number of ids in the system, despite lies by Byzantine processes. Furthermore, to ensure that the AA outputs preserve the initial ordering, we propose a validation scheme that does not require additional messages, and thus do not have the overhead of\,\cite{crash-to-byz1,crash-to-byz2}.

We then analyze the resulting algorithm when $N$ is large compared to $t$. In the lines of the work for crash-faults reported in~\cite{sirocco2012}, we show that the AA-based approximation phase, and thus our algorithm, requires only a constant number of steps to converge when $N>t^2+2t$. Interestingly, in this case it also achieves tight namespace of size $N$, because our id selection scheme ensures that Byzantine nodes are not able to introduce more than $t$ identifiers.

Even in the favorable case above, the number of communication steps can be an impairment for time constrained applications. Therefore, we then address the challenge of performing order-preserving renaming in as few communication steps as possible. We show that, if the number of faults is bounded by $N>2t^2+t$, it can be solved in just 2 steps. This is done by having processes exchange their initial ids, perform one echoing step, and then use the numbers of echoes to calculate a new name.

\subsection*{Contributions} To our knowledge, our work is the first to address the order-preserving renaming with Byzantine faults. Furthermore, our results also improve the existing work on non order-preserving renaming. Our main contributions can be summarized as follows (recall that $N$ is the number of processes and $t$ is an upper bound on the number of faults).
 
Our first contribution is an algorithm that performs order-preserving renaming with optimal fault tolerance of $N>3t$, has the same time and message complexity as the existing crash-tolerant solution~\cite{order-preserving-synch}, and is more efficient than the previous (\emph{non} order-preserving) algorithm for the Byzantine model. Additionally, our algorithm presents an improvement on the namespace size, $N+t-1$, compared to the previous result of $2N$ for non order-preserving renaming\cite{byzantine-renaming}. 

Our second contribution is to show that, if $N>t^2+2t$ our algorithm has constant step complexity and, interestingly, achieves optimal namespace of size $N$.

Our last contribution consists in a fast algorithm for $N > 2t^2+t$ that employs only $2$ communication steps and achieves the target namespace of size $N^2$. 

All algorithms presented in this paper are deterministic. 

\subsection*{Paper Organization}
The remainder of this paper is organized as follows. In Section~\ref{sec:model} we introduce the system model and formally define the problem addressed in this paper. In Section~\ref{sec:rw} we discuss the existing work. Section~\ref{sec:log} is dedicated to the order-preserving renaming algorithm for $N > 3t$. In Section~\ref{sec:constant}, we analyze our algorithm and show that it performs strong renaming within constant number of steps if $N>t^2+2t$. In section~\ref{sec:two-step}, we present a fast $2$-step renaming algorithm. Finally, Section~\ref{sec:conclusions} presents conclusions and outlines the directions for future work.

\section{Model and Problem Definition}\label{sec:model}
The processes are arranged in a synchronous network of an a priory known size $N$, in which each pair of processes is connected by a direct communication link. The communication between two processes is performed by message passing. The links of each process are labeled by $1, \cdots, N$, where the links $1,\cdots, N-1$ are to the remaining processes and link $N$ is a self-loop. It is assumed that the processes know the label of the link through which any message is received.

Each correct process has a unique identifier,  originally only known to the process itself. Up to $t$ processes may be faulty and exhibit arbitrary behavior (these processes are named Byzantine processes); faulty processes may send messages with arbitrary content. Communication channels are assumed to be reliable.

The renaming problem can be formally defined by the following conditions~\cite{hagit90,survey-sm}:
\begin{itemize}
\item \emph{Validity:} Each new name is an integer in the range $[1 \cdots M]$. 
\item \emph{Termination:} Each correct process outputs a new name.
\item \emph{Uniqueness:} No two correct processes output the same new name. 
\end{itemize}

The particular case in which the size of the target namespace is equal to $N$ is called \emph{strong} renaming.

In this paper we are interested in the order-preserving variant of the renaming problem, which requires the following property:
\begin{itemize}
\item \emph{Order-preservation:} New names of correct processes preserve the order imposed by their original ids.
\end{itemize}
\section{Related Work}\label{sec:rw}

The renaming problem was originally introduced in\,\cite{hagit90} for the asynchronous message-passing model with crash failures. The authors present a non order-preserving algorithm that solves renaming with a target namespace of size $N+t$ and an order-preserving algorithm with a target namespace of size $2^t(N-t+1)-1$. Both bounds on the target namespace were shown to be optimal\,\cite{hagit90,lb-namespace}. 

Although renaming can be solved using consensus as a building block, there are several reasons to devise algorithms specifically for solving renaming. First, consensus based solutions cannot be used in asynchronous systems subject to faults\,\cite{flp}. On the other hand, as shown in\,\cite{hagit90}, the impossibility result does not apply to renaming (i.e., renaming is ``easier" than consensus). Furthermore, in synchronous systems, consensus based solutions are viable but may be inefficient. In these settings, consensus requires $\Omega\pp{N}$ steps\,\cite{consensus-lowerbound}, while renaming can be implemented in $\bigO\pp{\log N }$ communication steps\,\cite{renaming-synch}. In fact, renaming is considered the simplest non-trivial distributed computing task~\cite{renaming-synch}. It is therefore no surprise that, after the seminal work of~\cite{hagit90}, a significant research effort has been placed in devising efficient algorithms for the renaming task.  From this point, we limit our discussion to the results on renaming in the synchronous message passing model considered in this paper.

A strong renaming algorithm with crash-faults, with optimal time complexity of $\bigO\pp{\log N }$ is presented in\,\cite{renaming-synch} and works as follows. A process chooses a new name by selecting one bit at a time, starting with the high-order bit and working down to the low-order bit. In each step the processes exchange their ids and the intervals of the new name in which they are interested, then split the ids in half, choosing $0$ if their own id belongs to the first half, or $1$ otherwise, and repeat the procedure. \cite{renaming-synch} also proves the lower bound of $\Omega\pp{\log N}$ for the renaming task for $N>t$.

A strong order-preserving renaming algorithm with logarithmic step complexity has been presented in\,\cite{order-preserving-synch} which also addresses crash faults. In this algorithm, the processes run an \emph{approximate agreement}, (or AA), to choose their new name. Unlike in the case of exact agreement (or \emph{consensus}), in the task of AA processes start with arbitrary real values and output values within some bounded distance from each other (e.g.~\cite{approx-agreement,inexact-agreement}).  In the AA-based renaming, processes exchange their old ids, propose a new name for each id based on its ranks in the list of all identifiers they received. Due to crashes processes may have received different sets of identifiers and therefore may propose different names for the same process. These discrepancies are later reduced by AA instances for each identifier that brings the values within safe distance from each other. Recently, the authors of\,\cite{sirocco2012} made the algorithm presented in\,\cite{order-preserving-synch} early deciding, i.e. the complexity depends on $f$, the number of actual faults occurred in a given run. Thus the complexity is $\bigO(\log f)$. Interestingly, the authors observed that the algorithm can decide in constant number of steps if the number of actual faults is bounded by $N>2f^2$. This is because in that case the approximate agreement is able to converge in a constant number of iterations. 

Byzantine renaming in message-passing systems has been addressed in\,\cite{byzantine-renaming}, where the authors prove the lower bound of $N > 3t$ on the number of Byzantine failures for the renaming problem in this model. The paper, that addresses the \emph{non} order-preserving variant of the problem, adapts the automatic crash-to-Byzantine translation techniques introduced in\,\cite{crash-to-byz1,crash-to-byz2} to the crash-tolerant algorithm introduced in\,\cite{renaming-synch}. The algorithm tolerates $N > 3t$ Byzantine failures and runs in $\bigO\pp{\log N }$ steps. Due to previously highlighted limitations of the translation techniques of\,\cite{crash-to-byz1,crash-to-byz2}, the tight target namespace of the original protocol is not preserved. Namely, because Byzantine processes can announce different identifiers that correct processes are not able to recognize as faulty, in the resulting transformed algorithm the target namespace is increased to $2N$. 

\section{Order-preserving Byzantine Renaming}\label{sec:log}

In this section, we present what is, to the best of our knowledge, the first order-preserving renaming algorithm with Byzantine faults.  The algorithm requires $N>3t$, which is optimal \cite{byzantine-renaming}.  Semantically, our algorithm follows the structure of the order-preserving algorithm for the fail-stop model presented in~\cite{order-preserving-synch} employing the techniques of Byzantine approximate agreement (AA) introduced in~\cite{approx-agreement} with extensions that address two additional concerns. First, we limit the number of identifiers introduced by the faulty processes. Second, we ensure that, in spite of contradictory information sent by Byzantine participants, the instances of AA converge in a consistent way that will allow new names to preserve the initial ordering.

The algorithm, depicted in Alg.~\ref{alg:alg2}, uses two distinct phases, namely the id selection phase and the rank approximation phase, or \emph{voting}. The first phase takes a constant number of steps (namely, 4~steps) to limit the number of identifiers produced by faulty nodes. At the end of this phase, each node makes an estimate of the new id for each process. However, as we will see, these estimates are not precise enough to be order-preserving. The second phase of the algorithm takes a logarithmic number of steps and runs, in parallel, multiple coordinated Byzantine-tolerant approximate agreements on those estimates. We denote each step as a \emph{voting step}. By making appropriate validations on the votes of each process, we ensure that the agreement converges to values that are order preserving. In the following subsections we discuss each of these phases in detail.

\begin{algorithm}[t]
\caption{Order-preserving Byzantine Renaming}
\label{alg:alg2}
{
\scriptsize
\begin{tabbing}
xxx,\=xxxx\=xxxx\=xxxx\=xxxx\=xxxx\=xxxx\=xxxx\kill
01 \> \textbf{Init:}\\
02 \> \>$\delta = 1 + \frac{1}{3(N+t)}$;\\
\\
// id selection phase\\
\\
03 \> \textbf{In Step $r$ := $1$} \\
04 \> \> broadcast ($\langle \mbox{\textsc{ID}}, \mbox{$my\_id$} \rangle$);\\
05 \> \> \textbf{foreach} $id$: $\langle$ \textsc{Id}, $id \rangle$ received from a distinct link  \textbf{do}\\
06 \> \> \> Ids := Ids $ \cup $ $\{id\}$;\\
\\
07 \> \textbf{In Step $r$ := $2$} \\
08 \> \> \textbf{foreach} $id$ $\in$ Ids \textbf{do}\\
09 \> \> \> broadcast($\langle$\textsc{Echo}, $id \rangle$);\\
10 \> \> Ids := $\emptyset$;\\
11 \> \> \textbf{foreach} $id$: $\langle \mbox{\textsc{Echo}}, $id$ \rangle$ received from at least $N-t$ distinct\\
\> \>  links \textbf{do}\\
12 \> \> \> Ids := Ids $ \cup $ $\{id\}$;\\
\\
13 \> \textbf{In Step $r$ := $3$} \\
14 \> \> \textbf{foreach} $id$ $\in$ Ids \textbf{do}\\
15 \> \> \> broadcast($\langle$ \textsc{Ready}, $id \rangle$);\\
16 \> \> Ids := $\emptyset$;\\
17 \> \> \textbf{foreach} $id$: $\langle \mbox{\textsc{Ready}}, id \rangle$ received from at least $N-t$ distinct\\
\> \>  links \textbf{do}\\
18 \> \> \> timely := timely~$\cup$~$\{id\}$;\\
19 \> \> \textbf{foreach}  $id$: $\langle \mbox{\textsc{Ready}}, id \rangle$ received from at least $N-2t$ distinct\\
\> \> links and have not broadcast $\langle \mbox{\textsc{Ready}}, id \rangle$ \textbf{do} \\ 
20 \> \> \> Ids := Ids $ \cup $ $\{id\}$;\\
\\
21 \> \textbf{In Step $r$ := $4$} \\
22 \> \> \textbf{foreach} $id$ $\in$ Ids \textbf{do}\\
23 \> \> \> broadcast($\langle \mbox{\textsc{Ready}}, id \rangle$);\\
24 \> \> \textbf{foreach} $\langle \mbox{\textsc{Ready}}, id \rangle$ received from at least $N-t$ distinct\\
\> \> links \textbf{do}\\
25 \> \> \> accepted := accepted~$\cup$~$\{id\}$;\\
26 \> \> \textsc{sort} (accepted); \\
27 \> \> \textbf{foreach} $id$~$\in$ accepted \textbf{do} \\
28 \> \> \> ranks[$id$] := \textsc{rank}(accepted,$id$)$\times\delta$;\\
\\
// rank approximation phase \\
\\
29 \> \textbf{In Step $r$ := $5$ to} $3 \lceil \log t\rceil + 7$\\
30 \> \> votes := $\emptyset$;\\
31 \> \> broadcast ($\langle \mbox{\textsc{AA}}, \mbox{ranks} \rangle$);\\
32 \> \> \textbf{foreach} $\langle  \mbox{\textsc{AA}, R} \rangle$ received \textbf{do}\\
33 \> \> \> \textbf{if} \textsc{isValid} (timely, R) \textbf{then} \\
34 \> \> \> \> votes := votes~$\cup$~R;\\
35 \> \> ranks := \textsc{aproximate}(ranks, votes); // updates ``accepted'' multiset\\
\\
36 \> \> \textbf{if Step $r$ =} $3\lceil \log t\rceil +7$\\
37 \> \> \> \textbf{return} \textsc{round}(ranks[$my\_id$]);
\end{tabbing}
}
\end{algorithm}

\subsection{Id Selection Phase}

This phase is implemented in Steps~$1$ to $4$ of Alg.~\ref{alg:alg2}. The purpose of the id selection phase is to choose which identifiers should feed the rank approximation phase. Note that Byzantine processes can announce different ids to different peers; if their power is not constrained the number of ``fake'' ids may prevent correct processes from executing correctly. On the other end, the purpose of this phase is not to ensure that all correct processes select the exact same set of identifiers: that would be equivalent to solving consensus, which would have linear step complexity.  For convenience of exposition, ids belonging to correct processes are named as \emph{correct} ids. All other ids are referred to as \emph{Byzantine}, e.g. ids issued by Byzantine processes as their own or non-existent ids that Byzantine processes claim to have received from others.

The algorithm uses the following variables and functions: two different sets, namely \emph{timely} and \emph{accepted} that are used to collect ids; the variable \emph{ranks} is a sparse array where \emph{ranks}[$id$] stores a new name for each id in the \emph{accepted} set; the function $\mbox{\textsc{sort}}(set)$ orders the entries in $set$; finally, the function $\mbox{\textsc{rank}}(set,v)$ returns a position of value $v$ in the ordered set $set$.

At the end of this phase,  the following properties are ensured on the \emph{timely} and \emph{accepted} sets:

$\bullet$ The \emph{timely}$_p$ at every correct process $p$, includes \emph{all} correct ids;

$\bullet$ The \emph{accepted}$_p$ at every correct $p$ includes at most $N+t-1$ values in total;

$\bullet$ The \emph{accepted}$_p$ at every correct $p$ is such that:
$$\bigcup_{q: q~is~correct} \mbox{timely}_q \subseteq \mbox{accepted}_p,$$
i.e., if one value is considered timely by some correct process, this value is for sure included in the \emph{accepted} set by every other correct process (but not necessarily considered timely).

In detail, this phase of the algorithm works as follows. In Step~1, each correct process broadcasts its identifier in an \textsc{Id} message. In Step~2, processes echo the ids they have received in the previous step (\textsc{Echo} messages). Only ids that have been echoed at least $N-t$ times are considered for the following steps. This effectively limits the number of Byzantine ids.  Also, since all correct ids are echoed by the correct processes, all correct ids are taken to the next steps. Ids that satisfy the previous condition are broadcast in a \textsc{Ready} message in Step~3 and all ids for which at least $N-t$ \textsc{Ready} messages have been issued are added to the $timely$ set. A process that did not broadcast \textsc{Ready} for a given id in Step~3, but observes at least $N-2t$ \textsc{Ready} messages for that id, also broadcasts \textsc{Ready} for that id in Step~$4$. Then, all \textsc{Ready} messages from Steps~3 and~4 are accounted, and all ids for which at least $N-t$ \textsc{Ready} messages have been produced are added to the \emph{accepted} set. 

The \textsc{Echo} and \textsc{Ready} messages used here are similar to the control messages exchanged in the reliable broadcast algorithm of~\cite{BrachaToueg}, with the difference that here the processes terminate in $4$ steps, which is sufficient to guarantee the required properties for the \emph{timely} and \emph{accepted} sets. As mentioned previously, reliable broadcast algorithms of~\cite{BrachaToueg} require each node to know the identity of a sender. Therefore, if the ids are not known \emph{a priori} and all processes are broadcasting at the same time, Byzantine participants can collude such that more than $t$ messages issued by Byzantine nodes are delivered by the correct processes. In fact, any message received in the first step by at least $N-2t$ correct nodes can be delivered by a correct process. Therefore, in our id selection, the size of the \emph{accepted} set at a correct process is bounded by $N+t-1$. Note also that Byzantine processes may use correct ids as their own; this has no effect on the execution: since \emph{timely} and \emph{accepted} are sets, duplicate identifiers are discarded.

At the end of the id selection phase, each process sorts its \emph{accepted} set, and assigns a new name to each of these processes (including itself), which is the rank of that id in the sorted set  stretched by the factor $\delta = 1 + \frac{1}{3(N+t)}$. This factor is large enough to prevent names from clashing due to small disagreement errors in the approximate agreement, as we explain below.  The purpose of the second phase is to iteratively execute approximate agreement until the ranks calculated by each correct process are within safe distance. 

\subsection{Approximation Phase}

\begin{algorithm}[t]
\caption{Procedure \textsc{isValid}}
\label{alg:isvalid}
{
\scriptsize
\begin{tabbing}
xxx,\=xxxx\=xxxx\=xxxx\=xxxx\=xxxx\=xxxx\=xxxx\=xxxx\kill
01 \> \textbf{Function} \textsc{isValid} (timely, ranks) \textbf{returns} boolean \textbf{is} \\
02 \> \> \textbf{foreach} $id$, $id'$~$\in$~timely \textbf{such that} $id < id'$  \textbf{do}\\
03 \> \> \> \textbf{if} $id \notin$ ranks \textbf{or} $id'\notin$ ranks \textbf{or} ranks[$id'$]$-$ranks[$id$]$ < \delta$ \textbf{then}\\
04 \> \> \> \> \textbf{return} false;\\
05 \> \> \textbf{return} true;
\end{tabbing}
}
\end{algorithm}

\begin{algorithm}[t]
\caption{Procedure \textsc{approximate}}
\label{alg:approx}
{
\scriptsize
\begin{tabbing}
xxx,\=xxxx\=xxxx\=xxxx\=xxxx\=xxxx\=xxxx\=xxxx\=xxxx\kill
01 \> \textbf{Function} \textsc{approximate} (my\_ranks, all\_ranks) \textbf{returns} array of ranks \textbf{is} \\
02 \> \> new\_ranks $:=$ $\emptyset$;\\
\> \> \\
03 \> \> \textbf{foreach} $id$~$\in$ accepted \textbf{do} \\
04 \> \> \> votes[$id$] := $\emptyset$;\\
05 \> \> \> \textbf{foreach} R~$\in$ all\_ranks \textbf{do} \\
06 \> \> \> \> \textbf{if} $id$~$\in$~R \textbf{then}\\
07 \> \> \> \> \> votes[$id$] $:=$ votes[$id$]~$\sqcup$~R[$id$];\\
08 \> \>  accepted := \{$id$ $\in$ accepted : $|$votes[$id$]$|\geq N-t$\};\\
\\
09 \> \> \textbf{foreach} $id$~$\in$ accepted \textbf{do} \\
10 \> \> \> \textbf{for} $|$votes[$id$]$|$ $+$ 1 \textbf{to} N \textbf{do} //fill missing votes with valid vote\\
11 \> \> \> \>  votes[$id$] := votes[$id$]~$\sqcup$~my\_ranks[$id$];\\
12 \> \> \> \textbf{for} 1 \textbf{to} t \textbf{do} // remove $t$ extreme values\\
13 \> \> \> \> votes[$id$] := votes[$id$]~$\setminus$~$\{\mbox{\textsc{max}(votes[$id$])}\}$;\\
14  \> \> \> \> votes[$id$] := votes[$id$]~$\setminus$~$\{\mbox{\textsc{min}(votes[$id$])}\}$;\\
15 \> \> \> $\mbox{\textsc{sort}(votes[$id$])}$; \\
16 \> \> \> new\_ranks[$id$] $:=$ \textsc{avg}(\textsc{select}$_t$(votes[$id$]);\\
17 \> \> \textbf{return} new\_ranks;
\end{tabbing}
}
\end{algorithm}

The approximation phase, or \emph{voting}, starts in Step~5 and takes logarithmic number of steps to converge. This phase is based on the Byzantine-tolerant AA algorithm of~\cite{approx-agreement}.  The AA algorithm guarantees that, in spite of contradictory inputs from Byzantine processes, the processes output values within a bounded error. Moreover, it guarantees that the outputs are within the range of values issued by the correct processes. Unfortunately the ranks calculated in the id selection phase may not preserve the correct ordering. As a result, the ranges of the correct inputs may overlap. Without any additional care, AA may converge to values that are not order-preserving.

The verification function depicted in Alg.~\ref{alg:isvalid} aims at ensuring that the approximation is performed in the way consistent with the ordering of the original ids. The function \textsc{isValid} takes as input the \emph{timely} set of the local process and a \emph{ranks} array received from some other process. It makes two tests to check if the votes from the remote process are consistent. First, the votes must include a vote for each id in \emph{timely} (we remind that if $p$ and $q$ are correct processes, then \emph{timely}$_p$~$\subseteq$~\emph{accepted}$_q$, thus any vote that does not satisfy this invariant may be discarded as faulty). Second, it ensures that the new rankings for these ids appear in the correct order separated by the minimum safety margin of $\delta$. Note that a Byzantine process may send different votes to different processes and both can still be considered valid. However, the presented validity conditions are sufficient to ensure that the approximation of the validated values will be done in a consistent way.

In addition to the variables and functions introduced before, the second phase of our algorithm also needs the following data structures and auxiliary functions: the variable $R$ is a set of \emph{ranks} arrays; the function $\mbox{\textsc{Round}}(x)$ returns the integral value nearest to $x$; finally, the function $\mbox{\textsc{select}}_k(set)$ returns a choice of values from a set. These values are chosen to maximize the convergence rate of the approximate agreement. Later in the text we describe what is the most appropriate choice function.

In detail, each voting step works as follows. Processes exchange the values in their \emph{ranks} array. Each array received from a remote process is first validated as described earlier. If the array is considered valid, it is accepted. Votes are then processed by the function \textsc{approximate}, depicted in Alg.~\ref{alg:approx}. In this function, each process computes a new rank for each id in the \emph{accepted} set as follows. It first collects all votes received for a given id into a multiset, (a multiset is a set that allows repetitions). If for some id in \emph{accepted}, less than $N-t$ votes are received, this id is discarded (by construction, this never happens to an id that has been considered timely by some correct process).  For the remaining ids, if the number of votes is less than $N$, process fills the multiset by including copies of its own value (intuitively, local values are always valid). Then, the resulting multiset of $N$ votes is sorted and the $t$ lower values and the $t$ higher values are discarded. Finally, function \textsc{select}$_t$ is used to pick a subset from the remaining values that is averaged to compute the new vote for that id. This function returns a multiset consisting of each $(it+1)$th element of the $set$ (which is an ordered multiset), where $0 \leq i < \lfloor \frac{\abs{set}}{t} \rfloor$. In other words, $\mbox{\textsc{select$_t$}}(set)$ returns a multiset consisting of the smallest and each $t$th element after it. This choice of \textsc{select}$_t$ is the same as in the approximate agreement algorithm of~\cite{approx-agreement}, which guarantees the convergence rate of $\sigma_t = \lfloor\frac{N-2t}{t}\rfloor +1$ where $\sigma_t$ is a number of elements returned by \textsc{select}$_t$ . 

After executing $3 \log t +7$ approximation steps, the new name is chosen as the rounded value of rank[$my\_id$]. The stretch factor of $\delta$ applied to the inputs and the validation procedure ensure that the ranks converge preserving a distance of slightly more than $1$, which prevents the rounded ranks from clashing in spite of a possible approximation error.

\subsection{Correctness}
Complete proofs are provided in Appendix~\ref{sec:app-a}.

We start by stating that any id in $timely$ at some correct process, is necessarily included in $accepted$ of any other correct process.
\begin{lemma}\label{thm:timely-all} For any $id$ such that  $id \in timely_p$ at some correct $p$, then $id \in accepted_q$ at any correct $q$.
\end{lemma}

The following lemma states that all correct ids are included in $timely$ sets of all correct processes.
\begin{lemma}\label{thm:correct-timely} If $id$ belongs to some correct $p$, then $id \in timely_q$ at any correct $q$.
\end{lemma}

As discussed earlier, Byzantine processes can generate more than $t$ identifiers, none of which recognized as faulty by the correct processes. The following lemma bounds the total number of ids added into $accepted$ by the correct processes.
\begin{lemma}\label{thm:init-namespace}At the end of Step~$4$, at each correct process \[\abs{accepted} \leq  N+ \lfloor \frac{t^2}{N-2t} \rfloor.\]
\end{lemma}

We then show that correct processes always issue valid votes.
\begin{lemma}\label{thm:valid} For any $r \geq 5$, if $ranks_p$ and $ranks_q$ are held by any two correct $p$ and $q$ in Step~$r$, then \[\textsc{isValid}(ranks_p,ranks_q)=true.\]
\end{lemma}
\begin{corollary}\label{thm:timely-never-excluded} If $id \in timely_p$ at some correct $p$, then its rank is updated in each approximation step by every correct process.
\end{corollary}
\begin{corollary}\label{thm:for-correct-id-dist-preserved} If $id < id'$ belong to two correct processes, then \[ranks_p[id]+ \delta \leq ranks_p[id'],\] at any correct $p$ in every Step $r \geq 4$.
\end{corollary}

We now need to bound the maximum discrepancy in the initial ranks for the same ids.
\begin{lemma}\label{thm:init-discrepancy} If $id \in timely_p$ for some correct $p$, then at the end of Step~$4$, \[\abs{ranks_p[id] - ranks_q[id]} \leq (t+ \lfloor \frac{t^2}{N-2t} \rfloor) \times \delta,\] where $ranks_q[id]$ is the rank of $id$ at some correct $q$.
\end{lemma}

Now it remains to show that each approximation step of Alg.~\ref{alg:approx} reduces the distance between the ranks by the factor $\sigma_t = \lfloor\frac{N-2t}{t}\rfloor +1$.
\begin{lemma}\label{thm:conv} Let $id \in timely_p$ at some correct $p$, and $\Delta_r$ denote the maximum distance between the correct ranks for $id$ in the beginning of Step~$r$. Then, at the end of Step~$r$, the distance between new correct ranks for this $id$ is within the range of $\frac{\Delta_r}{\sigma_t}$. Moreover, the new values are within the range of the old values belonging to correct processes.
\end{lemma}

We now calculate the number of iterations needed to reduce $\Delta_r$ to less than $\frac{1-\delta}{2}$.
\begin{lemma}\label{thm:final-dist} If $\Delta_5 \leq (2t-1) \times \delta$, then after $r = 3\lceil \log t \rceil +3$ iterations, the range of the values belonging to all correct processes is less than $\Delta_{r+4} < \frac{\delta-1}{2}$. 
\end{lemma}

Finally, we are ready to prove the main theorem.
\begin{theorem}\label{thm:final} Alg.~\ref{alg:alg2} implements order-preserving renaming for $N>3t$ with the target namespace of size $N+t-1$.
\end{theorem}
\begin{IEEEproof}

\textbf{Validity}. By Lemma~\ref{thm:init-namespace}, $\abs{accepted} \leq N + \lfloor \frac{t^2}{N-2t} \rfloor \leq N +t-1$, for $N>3t$. Therefore, the initial ranks are bounded by $(N+t-1)\times \delta$. Since by Lemma~\ref{thm:conv}, all correct processes output a value within the interval of the initial correct values,  the outputs of the correct processes are bounded by \textsc{round}$\pp{(N+t-1)\times \delta} = N+t-1$.

\textbf{Termination}. After $3 \lceil \log t \rceil + 7$ steps, every correct process outputs a value.

\textbf{Order-preserving}. By Lemmas~\ref{thm:correct-timely}, correct ids are always included in $timely$ sets and, by Corollary~\ref{thm:timely-never-excluded}, are updated in each step by every correct process. By Corollary~\ref{thm:for-correct-id-dist-preserved}, for any two correct $id$ and $id'$ such that $id < id'$, the distance between their rankings is lower bounded by $\delta$ in every step.
Since by Lemma~\ref{thm:final-dist}, after $3\lceil  \log t \rceil + 7$ steps, $\Delta_r<\frac{\delta-1}{2}$,
\mymathline{rank\pp{id} + \delta + \frac{1-\delta}{2} < rank\pp{id'} - \frac{1-\delta}{2}.}
Hence,
\textsc{Round}$\pp{ranks[id])}<$ \textsc{Round}$\pp{ranks[id']}.$
\end{IEEEproof}

\subsection{Complexity Analysis}
The step complexity of Alg.~\ref{alg:alg2} is $3\lceil \log t \rceil + 7$. 
In each step, the processes employ all-to-all communication. Hence, the total message complexity is $\bigO \pp{N^2 \log t}$. Since in each communication the processes exchange arrays of at most $N+t-1$ original ids and their ranks, the message size is bounded by $\bigO\pp{(N+t-1)\pp{ \log N_{max} + \log N}}$ bits.

\section{Constant Time Renaming}\label{sec:constant}

An interesting property of Alg.~\ref{alg:alg2} is that it performs strong renaming, i.e. renaming with the target namespace of $N$, within constant number of steps if $N>t^2+2t$. The optimal namespace is due to the fact that Byzantine processes are not able to introduce any additional identifiers in our id selection scheme. The constant step complexity is due to the fast convergence property of the Byzantine AA. Similar argument was used by the authors of~\cite{sirocco2012} to prove the constant step complexity of the crash-tolerant algorithm presented in~\cite{order-preserving-synch} when the number of crashes is bounded by $N > 2t^2$. 
This result is formalized below. Proofs are provided in Appendix~\ref{sec:app-b}.
\begin{lemma}\label{thm:namespace-const}
For $N>t^2+2t$, Alg.~\ref{alg:alg2} achieves the target namespace of size $N$.
\end{lemma}
\begin{lemma}\label{thm:conv-const}
After $4$ approximation steps, the values held by the correct processes are within the distance of less than $\frac{\delta-1}{2} = \frac{1}{6(N+t)}$.
\end{lemma}

Therefore, if we change the code of Alg.~\ref{alg:alg2} to run only $4$ approximation steps (Line 29), as a result the algorithm has the complexity of 8 steps.
\begin{theorem}\label{thm:constant}  Alg.~\ref{alg:alg2} implements strong order-preserving renaming in $\bigO(1)$ steps if $N>t^2+2t$.
\end{theorem}

\section{2-Step Renaming Algorithm}\label{sec:two-step}

In the previous section we have shown that Alg.~\ref{alg:alg2} has constant step complexity for $N>t^2+2t$. This is an interesting result from the asymptotic point of view, specially considering that the resulting name space is optimal. Still, from the practical point of view, the number of communication steps can still be an impairment for time constrained applications (the number of steps of  Alg.~\ref{alg:alg2} is exactly 8). Therefore, in this section we are interested in performing renaming in as few communication steps as possible.
Interestingly, we show that order-preserving renaming in face of Byzantine processes can be solved in just 2 communication steps in the case $N>2t^2+t$, by relaxing the target namespace to $N^2$. Obviously, in just 2 communication steps, it is impossible to perform iterative approximate agreement. In fact, our algorithm is simply based on counting echoes that are filtered by a validity check. 

The algorithm is depicted in Alg.~\ref{alg:alg1}. The main idea of the algorithm is having each process initially announce its ids to all other processes; then, echo all the ids received in the first step, and finally having each correct process calculate its new name by ordering all the received ids, and calculating \emph{offsets}, i.e. spacings between two consecutive names, according to the number of echoes received for each id.  Byzantine processes may opt not to echo the ids or even send contradictory information to different processes. Therefore, correct processes may receive different sets of ids as well as different numbers of echoes for each ids. The key to the algorithm is to compute the offsets in such a way that the new names chosen by the correct processes will hold the order-preserving property, despite the potentially inconsistent input sets of echoes.

\begin{algorithm}[t]
\caption{2-step Order-preserving Byzantine Renaming for $N>2t^2+t$}
\label{alg:alg1}
{
\scriptsize
\begin{tabbing}
xxx,\=xxxx\=xxxx\=xxxx\=xxxx\=xxxx\=xxxx\=xxx\=xxx\kill
01 \> \textbf{Init:}\\
02 \> \> \textbf{foreach} $\mbox{lnk} \in \{ 1, \cdots, N\}$  linkid$[lnk]:= \perp$; \\
03 \> \> timely := accepted := $\emptyset$;\\
04 \> \> \textbf{forall} $id$ \textbf{do} counter[$id$] := 0; // init sparse array with zeros \\
\\
05 \> \textbf{In Step $r$ := $1$} \\
06 \> \> broadcast ($\langle \mbox{\textsc{ID}}, my\_id \rangle$);\\
08 \> \> \textbf{foreach} $id$: $\langle \mbox{\textsc{Id}}, id \rangle$ received from a distinct link \emph{lnk} \textbf{do}\\
09 \> \> \> linkid$[lnk]$ := $id$;\\
10 \> \> \> timely := timely~$\cup$~$\{id\}$;\\
\\
11\> \textbf{In Step $r$ := $2$} \\
12 \> \> broadcast ($\langle \mbox{\textsc{MultiEcho}}, \mbox{timely} \rangle$);\\
 \> \> // count echoes\\
13 \> \> \textbf{foreach} $id$: $\langle \mbox{\textsc{MultiEcho}}, \mbox{ids} \rangle$ received from a distinct \\
\> \> link \emph{lnk} \textbf{do}\\
14 \> \> \> \textbf{if} \textsc{isValid} (\emph{lnk, ids}) \textbf{then} \\
15 \> \> \> \> \textbf{foreach} $id \in$ ids \textbf{do}\\
16 \> \> \> \> \> accepted:= accepted~$\cup$~$\{id\}$;\\
17 \> \> \> \> \> counter[$id$] := counter[$id$] +1;\\
\> \> // compute new names \\
18 \>\> \textsc{sort} (accepted); \\
19 \> \> accum\_offset := $0$; \\
20 \> \> \textbf{for} $id$ := \textsc{first}(accepted) \textbf{to} \textsc{last}(accepted) \textbf{do}  \\
21 \> \> \> accum\_offset := accum\_offset + \textsc{min} (counter[$id$], $N-t$); \\
22 \> \> \> newid[$id$] : = accum\_offset; \\
23 \> \> \textbf{return} newid[$my\_id$]\\
\\
01 \> \textbf{Function} \textsc{isValid} (lnk, ids) \textbf{returns} boolean \textbf{is} \\
02 \> \> \textbf{return} linkid$[\mbox{lnk}] \neq \perp) \land$ $(\abs{\mbox{ids}} \leq N) \land $ $(\abs{ \mbox{timely}\cap \mbox{ids}} \geq N-t$)
\end{tabbing}
}
\end{algorithm}

As the previous algorithms, Alg.~\ref{alg:alg1} also uses a \emph{timely} and an \emph{accepted} set of ids. However, in this algorithm, all ids broadcast in Step~$1$ are considered timely and all ids echoed in Step~$2$, that pass a basic validity test, are accepted. The validity test, captured by function \textsc{isValid}, limits the power of Byzantine processes as follows: first it only accepts echo messages from processes that have sent their id in Step~$1$, then it does not accept a \textsc{MultiEcho} message that has more than $N$ ids, finally that the incoming \textsc{MultiEcho} has at least $N-t$ ids in common with the timely set of the recipient (note that if the sender and recipient of a \textsc{MultiEcho} are correct, they both have at least the $N-t$ correct processes in their timely set). Also, for each accepted id, the algorithm counts how many times that id has been echoed by all processes (again, correct ids are guaranteed to be echoed at least $N-t$ times).

After all echo messages have been processed, processes are ready to choose new names. The offset for each known id is simply the value of $\mbox{\textsc{min}}(counter,N-t)$ (Line 21). The adjustment to $N-t$ will guarantee that these offsets for the correct ids are always the same. This prevents Byzantine processes from introducing an additional error linear in the number of correct processes by choosing to echo correct ids for some processes but not others.  Finally, the new name of the process is produced by summing the offsets of all ids up to, and including, the id of the process executing the algorithm. 
The algorithm also stores the (locally estimated) values of new names for other processes; this is not required in practice and is done here only for clarity of the proofs. 

\subsection{Correctness}
Proofs are provided in Appendix~\ref{sec:app-c}.

Let $\Delta$ denote the maximum possible discrepancy between the new names for some correct id.
\begin{lemma}\label{thm:deviation}  $\Delta \leq 2t^2.$
\end{lemma}

We now establish the minimum offset of any correct id.
\begin{lemma}\label{thm:init-dist}  Let $id$ and $id'$ be two correct identifiers. If $id' < id$, then $newid_p[id'] + (N-t) \leq newid_p[id]$ at some correct $p$.
\end{lemma}

We are ready to prove main theorem.
\begin{theorem}\label{thm:2step}   Alg.~\ref{alg:alg1} implements order-preserving renaming for $N>2t^2+t$ with the target namespace of size $N^2$.
\end{theorem}
\begin{IEEEproof}
\textbf{Validity}. The total number of echoed ids accepted by each correct process in Step~$2$ is at most $N^2$. Therefore, the correct processes output an integer value within the range $[1,\cdots,N^2]$, meaning that Alg.~\ref{alg:alg1} satisfies the validity property.

\textbf{Termination}. After $2$ steps, every correct process outputs a value.

\textbf{Order-preserving}. Consider two correct processes $p$ and $q$ with initial identifiers $id$ and $id'$, such that $id<id'$. By Lemma~\ref{thm:init-dist}, $newid_p[id] + N-t \leq newid_p[id']$. Since by Lemma~\ref{thm:deviation}, $\Delta \leq 2t^2$, meaning that $newid_p[id'] - 2t^2 \leq newid_q[id']$. Since $N>2t^2+t$, \[newid_p[id]+N-t-2t^2 < newid_q[id'].\]
\end{IEEEproof}

\subsection{Complexity Analysis}
Alg.~\ref{alg:alg1} consists of $2$ communication steps. Since in each step, processes employ all-to-all communication, the total message complexity is $2 N^2$. In Step~$2$, the processes exchange vectors of all ids they received in Step~$1$. Therefore, the message size is bounded by $\bigO\pp{N \log N_{max}}$ bits.

\section{Conclusions}\label{sec:conclusions}

This paper addresses for the first time the problem of order-preserving renaming in synchronous systems subject to Byzantine faults. However, our contributions also improve the existing results on non order-preserving renaming in this model.

Our first algorithm performs order-preserving renaming with optimal fault tolerance of $N>3t$, has the same time and message complexity as the existing crash-tolerant solution~\cite{order-preserving-synch} and is more efficient than the previous (non order-preserving) algorithm for the Byzantine model. Additionally, our algorithm presents an improvement on the namespace size compared to the previous result of~\cite{byzantine-renaming} and even achieves optimal namespace size for $N>t^2+2t$. It remains an open question whether it is possible to achieve tight namespace and optimal fault tolerance without using consensus. 

On the other hand, when the number of Byzantine faults is on the order of $\sqrt N$, we have shown that renaming can be performed in constant time both by using approximate agreement and with a simple echo-scheme. This bound on the number of faults asymptotically matches the existing results for the crash-fault model~\cite{sirocco2012}. Another open question is whether this bound is optimal or better fault tolerance can be achieved in constant time.

\fakeparagraph{Acknowledgments:} The authors are thankful to Marcos K. Aguilera for his comments on an earlier version of this document.

\bibliographystyle{plain}
\bibliography{bibliography}

\appendices
{

\section{}\label{sec:app-a}

\begin{IEEEproof}[Proof of Lemma~\ref{thm:timely-all}]
Assume by contradiction, $id \notin accepted_q$ at some correct $q$. This is only possible if $q$ has not received $N-t$ $\langle \mbox{\textsc{Ready}}, \mbox{id} \rangle$ messages in Steps 3 and 4. But if $p$ added $id$ into $timely$, it means that it has received at least $N-t$ $\langle \mbox{\textsc{Ready}}, \mbox{id} \rangle$ messages, $N-2t$ of which must have been sent by the correct processes in Step~$3$ (Lines 17-18 of Alg.~\ref{alg:alg2}), therefore the correct processes that have not issued $\langle \mbox{\textsc{Ready}}, \mbox{id} \rangle$  in Step~$3$ will do so in Step~$4$ (Lines 22-23). It means that all correct processes issue $\langle \mbox{\textsc{Ready}}, \mbox{id} \rangle$  by Step~$4$, which leads to a contradiction.
\end{IEEEproof}

\begin{IEEEproof}[Proof of Lemma~\ref{thm:correct-timely}]
Assume by contradiction, $id \notin timely_q$ for some correct $q$. This means that $q$ has not received $N-t$ $\langle \mbox{\textsc{Ready}}, \mbox{id} \rangle$ in Step~$4$. This is only possible if some correct process has not issued $\langle \mbox{\textsc{Ready}}, \mbox{id} \rangle$, which in turn is because it has not received $N-t$ $\langle \mbox{\textsc{Echo}}, \mbox{id} \rangle$ in Step~$2$. This also is only possible if $id$ was not received by some correct process in Step~$1$. However, since $p$ is correct, $p$ sent $id$ to all correct processes in Step~$1$. Contradiction.
\end{IEEEproof}

The following lemma will be used to calculate the maximum number of identifiers that Byzantine processes are able to produce.
\begin{lemma}\label{thm:accepted} If $id \in accepted_p$ at some correct process $p$, then at least $N-2t$ correct processes received $id$ in Step~$1$.
\end{lemma}
\begin{IEEEproof}[Proof of Lemma~\ref{thm:accepted}]
 If $id \in accepted$, then $p$ has received at least $N-t$ $\langle\mbox{\textsc{Ready}},id\rangle$ messages from which at least $N-2t$ must have been issued by the correct processes. From all $\langle\mbox{\textsc{Ready}},id\rangle$ issued by the correct processes, at least one is sent in Step~$3$ (Line 19-20). This means that some correct process received at least $N-t$ $\langle\mbox{\textsc{Echo}},id\rangle$ messages in Step~$2$, $N-2t$ of which must have come from the correct processes.
\end{IEEEproof}

\begin{IEEEproof} [Proof of Lemma~\ref{thm:init-namespace}]
By Lemma~\ref{thm:correct-timely}, all $N-t$ correct ids are in $timely$, therefore also in $accepted$.
It remains to calculate the maximum number of Byzantine ids that can be in $accepted$. By Lemma~\ref{thm:accepted}, each $id \in accepted$ must have been broadcast in Step~$2$ by at least $N-2t$ correct processes. This means that from the total of at most $t (N-t)$ identifiers broadcast by the Byzantine processes in Step~$1$, $\lfloor \frac{t (N-t)}{N-2t} \rfloor =  t+ \lfloor \frac{t^2}{N-2t} \rfloor$ can be in $accepted$ at any correct process at the end of Step~$4$. 
\end{IEEEproof}

The following lemma is auxiliary and states that if we construct two multisets by adding pairwise values separated by some given distance from each other, then after we order the multisets, the entries on the corresponding indexes still preserve this distance.

\begin{lemma}\label{thm:dist}  Let $U$ and $W$ be two ordered multisets with $k$ elements each, created by adding $k$ pairs of elements $a,pair(a)$ into $U, W$ respectively, such that $a + \delta \leq pair(a)$. Then,  for any $1 \leq i \leq k$, $u_i + \delta \leq w_i$. 
\end{lemma}
\begin{IEEEproof} 
We first show that the inequality holds for the first elements in the ordered multisets, i.e. \myequationa{eq00}{u_1}{ + \delta \leq w_1.}

Since $w_1$ is the smallest in $W$, $w_1\leq pair(u_1)$. If $w_1= pair(u_1)$, then \eqref{eq00} follows. If $w_1 < pair({u_1})$, there exists $u_i$ such that $w_1 = pair({u_i})$. Since $u_1$ is the smallest in $U$, $u_1 + \delta \leq u_i + \delta \leq w_1$, as claimed. 

Now, by making $pair({u_1})$ a new pair of $u_i$, the same argument is used to iteratively prove \eqref{eq00} for $U = U \setminus \{u_1\}$ and $W = W \setminus \{w_1\}$ until $U$ and $W$ are empty. Therefore, $1 \leq i \leq k$, $u_i + \delta \leq w_i$, as needed.
\end{IEEEproof}

The following lemma shows that during the approximation procedure, the distance between the ranks of two ids included in the $timely$ set of some correct process maintains at least $\delta$.

\begin{lemma}\label{thm:min-dist} If for some ids $id, id' \in timely$, at the beginning of Step~$r$, $ranks[id] + \delta \leq ranks[id']$ and $\abs{votes[id]} ,\abs{votes[id']} \geq N-t$, then at the end of Step~$r$, $ranks[id] + \delta \leq ranks[id']$.
\end{lemma}
\begin{IEEEproof} 
Since $id, id' \in timely$, all votes accepted in Line 25 must contain new ranks for both $id$ and $id'$ spaced by at least $\delta$. Hence, $\abs{votes[id]} = \abs{votes[id']}$.
 
If there are less than $N$ entries in each set, the $ranks[id]$ and $ranks[id']$ will be added respectively such that both sets have exactly $N$ entries (Lines 10-11 of Alg.~\ref{alg:approx}), (by assumption, the added values also preserve the distance of at least $\delta$).

Now, assume $U,W$ are multisets resulted from ordering $votes[id]$ and $votes[id']$ respectively. By Lemma~\ref{thm:dist}, for any $1\leq i \leq N$, $u_i + \delta \leq w_i$. Hence, after deleting from $U$ and $W$, $t$ smallest and $t$ largest entries (Line 13-14 of Alg.~\ref{alg:approx}), it still holds that $1\leq i \leq N-2t$, $u_i + \delta \leq w_i$.
The distance between the new values (calculated in Line 16) is given by,
\begin{eqnarray}
&&\mbox{\textsc{avg(select$_t$}}(W)) - \mbox{\textsc{avg(select$_t$}}(U)) \nonumber \\ 
&& \geq \quad \frac{\mbox{\textsc{sum(select$_t$}}(U))+ t\delta}{t} - \frac{\mbox{\textsc{sum(select$_t$}}(U))}{t} \nonumber \\
&& = \quad \delta. \nonumber 
\end{eqnarray}
\end{IEEEproof}

\begin{IEEEproof}[Proof of Lemma~\ref{thm:valid}]
$\textsc{isValid}(ranks_p,ranks_q)$ checks if the distance between the ranks of all elements in $timely_p$ is at least $\delta$. By Lemma~\ref{thm:timely-all}, $timely_q \subseteq accepted_p$. Therefore, if the entries in $ranks_p$ preserve the distance of least $\delta$, for any $id$ such that $id \in \bigcup_{q:~q~is~correct} timely_q$, in Step~$r$, then $\textsc{isValid}(ranks_p,ranks_q)$.

We now show by induction on $r$ that the distance between the ranks of ids in $timely_p$ is preserved at least $\delta$ by all correct processes in any Step~$r \geq 5$. For the base case of $r=5$, recall that $p$ constructs the initial ranks in such a way that all ranks for the $accepted$ set are spaced by at least $\delta$ (Line 28 of Alg.~\ref{alg:alg2}), therefore $\textsc{isValid}(ranks_p,ranks_q)=true$. 

For the induction step, assume that, for the $rank$ held by $p$ in Step~$r$, $\textsc{isValid}(ranks_p,ranks_q)=true$. Therefore, for each element in $timely$ each correct process will receive at least $N-t$ valid votes. And since by assumption, the correct votes are valid in Step~$r$ and by Lemma~\ref{thm:correct-timely} each correct vote contains new ranks for all ids in $timely_p$, $p$ will update their values in Line 35 of Alg~\ref{alg:alg2} and, by Lemma~\ref{thm:min-dist}, the new ranks calculated by each correct process at the end of Step~$r$ preserve the necessary distance at least $\delta$. Therefore, $\textsc{isValid}(rank_p,rank_q)=true$ in $r+1$.
\end{IEEEproof}

\begin{IEEEproof}[Proof of Lemma~\ref{thm:init-discrepancy}]
 By assumption, $id \in timely_p$, therefore, by Lemma~\ref{thm:timely-all}, $id \in accepted_q$. Also, by Lemma~\ref{thm:correct-timely}, all correct ids are in $timely_p$ and $timely_q$ and therefore in $accepted$ at each correct process. Hence, $\abs{accepted_p \cap accepted_q} \geq N-t$. On the other hand, by Lemma~\ref{thm:init-namespace}, all correct processes have $\abs{accepted} \leq N+t-1$. Hence, the initial ranks calculated in Line 28 of Alg~\ref{alg:alg2} of each common element of $accepted_p$ and $accepted_q$ differs by at most $(2t-1) \times \delta$. 
\end{IEEEproof}

\begin{IEEEproof}[Proof of Lemma~\ref{thm:conv}]
Since $id \in timely_p$, then by Lemma~\ref{thm:valid} and Corollary~\ref{thm:timely-never-excluded}, $votes_p[id]$ and $votes_q[id]$ have at least $N-t$ entries from the correct processes, therefore after executing Lines 12-14 of Alg.~\ref{alg:approx} both multisets have exactly $N$ entries.

Let $C$ be the multiset of ranks of $id$ issued by all correct processes in Alg.~\ref{alg:alg2}, in Step~$r$. Note that $C \subseteq votes_p[id], votes_q[id]$. 

Let $A, B$ be ordered multisets resulting from deleting $t$ maximal values and $t$ minimal values from $votes_p[id]$ and $votes_q[id]$, respectively. Let $a_1 \leq \cdots \leq a_c$ be the elements of $\mbox{\textsc{select$_t$}}(A)$ and $b_1 \leq \cdots \leq b_c$ be the elements of $\mbox{\textsc{select$_t$}}(B)$, where $c$ is the number of elements selected. Note that $c=\sigma_t$.

First, we need to show that, for $1 \leq i \leq c-1$, \myequationa{eq:x}{\mbox{\textsc{max}}(a_i,b_i)}{ \leq \mbox{\textsc{min}}(a_{i+1},b_{i+1}).}

It suffices to show that $a_i \leq b_{i+1}$, then by symmetric argument $b_i \leq a_{i+1}$. Suppose, by contradiction, that $a_i > b_{i+1}$. There are at least $t(i+1)+1$ elements in $B$ less than or equal to $b_{i+1}$. By our supposition, these elements are strictly less than $a_i$. However, there are at most $ti$ elements in $A$ strictly less than $a_i$. Therefore, at least $t(i+1) + 1 - ti = t +1$ elements in $B$, are not in $A$. However, since $\abs{votes_p[id] \cap votes_q[id]} \geq N-t$, it holds that $\abs{A \cap B} \geq N-t-2t$. Therefore, \[\abs{B-A} = \abs{ B - \pp{A \cap B}} \leq (N - 2t) - (N-3t) = t.\] Hence the contradiction and~\eqref{eq:x} follows.

We then use~\eqref{eq:x} to prove the lemma. The discrepancy between $ranks_p[id]$ and $ranks_q[id]$, which are updated in Line 16 of Alg.~\ref{alg:approx} at the end of Step~$r$, is given by,
\begin{eqnarray}
&&\abs{ \mbox{\textsc{avg(select$_t$}}(A)) - \mbox{\textsc{avg(select$_t$}}(B))}\nonumber\\
&& = \quad  \frac{1}{c}\abs{\pp{a_1 + \cdots + a_c} - \pp{b_1 + \cdots + b_c}}\nonumber\\
&& =  \quad \frac{1}{c} \abs{\sum_{i=1}^c (a_i-b_i)}\nonumber\\
&& \leq  \quad \frac{1}{c} \sum_{i=1}^c \abs{a_i-b_i}\nonumber\\
&& =  \quad \frac{1}{c}\sum_{i=1}^c \pp{\mbox{\textsc{max}}(a_i,b_i) - \mbox{\textsc{min}}(a_i,b_i)}, \label{eq2}
\end{eqnarray}
where the fourth line follows from triangular inequality.

Expanding the sum and successively applying~\eqref{eq:x},
 \begin{eqnarray}
&& \frac{1}{c}\sum_{i=1}^c \pp{\mbox{\textsc{max}}(a_i,b_i) - \mbox{\textsc{min}}(a_i,b_i)}\nonumber\\
&& = \quad \frac{1}{c} \pp{\mbox{\textsc{max}}(a_c,b_c) - \mbox{\textsc{min}}(a_c,b_c)} \nonumber\\
&& + \quad   \frac{1}{c}\sum_{i=1}^{c-1} \pp{\mbox{\textsc{max}}(a_i,b_i) - \mbox{\textsc{min}}(a_i,b_i)} \nonumber\\
&& \leq \quad \frac{1}{c}  \pp{\mbox{\textsc{max}}(a_c,b_c) - \mbox{\textsc{min}}(a_1,b_1)}. \label{eq3}
\end{eqnarray}
On the other hand, since we deleted $t$ extremal values from $votes_p[id]$ and $votes_q[id]$, it is true that $\mbox{\textsc{max}}(a_c,b_c) \leq \mbox{\textsc{max}}(C)$ and $\mbox{\textsc{min}}(a_1,b_1) \geq \mbox{\textsc{min}}(C)$. Therefore, the averages are within the interval of the input values belonging to the correct processes.
 
Moreover, from~\eqref{eq2} and~\eqref{eq3},
\begin{eqnarray}
&&\abs{ \mbox{\textsc{avg(select$_t$}}(A)) - \mbox{\textsc{avg(select$_t$}}(B))}\nonumber\\
&& \leq  \quad \frac{1}{c} \pp{\mbox{\textsc{max}}(C) - \mbox{\textsc{min}}(C)}\nonumber\\
&& = \quad  \frac{1}{\sigma_t}\Delta{_r}.\nonumber 
\end{eqnarray}
Hence, the lemma follows.
\end{IEEEproof}

\begin{IEEEproof}[Proof Lemma~\ref{thm:final-dist}]
By successive applications of Lemma~\ref{thm:conv},
\begin{eqnarray}
\Delta_{r+4}
&\leq&  \pp{\frac{1}{\sigma_t}}^{r} \Delta_5\nonumber\\
& < &  \pp{\frac{1}{2}}^{\lceil 3\log \pp{t}\rceil+3} 2t \times \pp{1+\frac{1}{N+t}}\nonumber \\
&<& \frac{1}{6\pp{N+t}}. \nonumber
\end{eqnarray}
\end{IEEEproof}

\section{}\label{sec:app-b}

\begin{IEEEproof}[Proof of Lemma~\ref{thm:namespace-const}]
By Lemma~\ref{thm:init-namespace}, the number of ids in the $accepted$ set of any correct process is at most $N + \lfloor \frac{t^2}{N-2t} \rfloor = N$. Due to the stretching factor of $\delta = 1 + \frac{1}{3(N+t)}$, the initial ranks are bounded by $N\times\delta$. Since by Lemma~\ref{thm:conv} the values returned by the approximation belong to the interval of the initial correct values, the rounded outputs will be at most \textsc{round}$\pp{N\times\delta} = N$.
\end{IEEEproof}

\begin{IEEEproof}[Proof of Lemma~\ref{thm:conv-const}]
By Lemma~\ref{thm:init-discrepancy}, the maximum discrepancy between the votes is at most $(t + \lfloor \frac{t^2}{N-2t} \rfloor)\times \delta = t \times \delta$. On the other hand, by Lemma~\ref{thm:conv}, the convergence rate of each approximation step is at least  $\sigma_t = \lfloor\frac{N-2t}{t}\rfloor +1 > \lfloor\frac{t^2}{t}\rfloor +1 = t+1$. Therefore, after $4$ convergence steps, the values of the correct processes are within
\mymathline{\frac{t \times \delta}{(t+1)^4} < \frac{1}{3t^3} < \frac{\delta-1}{2}.\vspace{-\baselineskip}}
\end{IEEEproof}

\section{}\label{sec:app-c}

\begin{IEEEproof}[Proof of Lemma~\ref{thm:deviation}]
For each echo message received in Step~$2$, a correct process compares the number of ids in common, that should be at least $N-t$ out of $N$ allowed per message (procedure \textsc{isValid}). Due to this sanity check, each Byzantine process can introduce only $2t$ Byzantine ids in an echo message: in the worst case, the Byzantine process includes $t$ Byzantine ids already known to the receiver and some additional $t$ arbitrary ids. Therefore, the total number of echoes of Byzantine ids received from the Byzantine processes by each correct process in Step~$2$, is at most $2t^2$.
\end{IEEEproof}
}

\begin{IEEEproof}[Proof of Lemma~\ref{thm:init-dist}]
Assume, by contradiction, that $newid_p[id] - newid_p[id'] < N-t$. This is only possible if $counter_p[id] < N-t$ (Line 21). This means that, in Step~2, $p$ received less than $N-t$ echoes of $id$. It can only happen if some correct process $p'$ did not echo $id$. This, in turn, is only possible if $p'$ did not receive $id$ in Step~$1$. But since $id$ is correct, it was sent to all the processes in Step~$2$. Contradiction.
\end{IEEEproof}

\end{document}

%% file: macros.tex
\newcommand{\abs}[1]{\left| #1 \right|}

\newcommand{\lp}{\left(}
\newcommand{\rp}{\right)}
\newcommand{\pp}[1]{\lp #1 \rp}

\newcommand{\bigO}{\mathcal{O}}

\newcommand{\mymathline}[1]
{
$$
#1 
$$
}

\newcommand{\myequationa}[3]
{
\begin{align}
#2
   & #3 \label{#1}
\end{align}
}